\documentclass[a4paper,11pt]{article}
\usepackage{jcappub} % for details on the use of the package, please see the JINST-author-manual
\usepackage{color}
\usepackage[dvipsnames]{xcolor}
\usepackage{lineno}
\usepackage{booktabs}
\usepackage{bbding}
\usepackage{tabularx}
\usepackage{subcaption}
\usepackage{multirow}
\usepackage{array}
\usepackage{float}

\usepackage{makecell} % in preamble
\arxivnumber{2505.13613}
\ETnumber{\, ET-0147A-25}
\title{Distinguishing Distance Duality breaking models using electromagnetic and gravitational waves measurements}

\author[a,b,c,d]{Chiara De Leo,}
\author[b,c]{Matteo Martinelli,}
\author[b,c]{Rocco D'Agostino,}
\author[e]{Giulia Gianfagna,}
\author[f,g]{C.~J.~A.~P. Martins}
\affiliation[a]{Sapienza University, Piazzale Aldo Moro, 2 - c/o Dipartimento di Fisica, Edificio E. Fermi, I-00185 Rome, Italy}
\affiliation[b]{INAF - Osservatorio Astronomico di Roma, via Frascati 33, 00078 Monte Porzio Catone, Italy}
\affiliation[c]{INFN - Sezione di Roma, Piazzale Aldo Moro,
2 - c/o Dipartimento di Fisica, Edificio G. Marconi, I-00185 Rome, Italy}
\affiliation[d]{Tor Vergata University, Via della Ricerca Scientifica, 1 - c/o Dipartimento di Fisica, I-00133 Rome, Italy}
\affiliation[e]{INAF - Istituto di Astrofisica e Planetologia Spaziali, via Fosso del Cavaliere 100, I-00133 Rome, Italy}
\affiliation[f]{Centro de Astrof\'isica da Universidade do Porto, Rua das Estrelas, 4150-762 Porto, Portugal}
\affiliation[g]{Instituto de Astrof\'isica e Ci\^encias do Espa\c{c}o, Universidade do Porto, Rua das Estrelas, 4150-762 Porto, Portugal}

\emailAdd{chiara.deleo@uniroma1.it}
\emailAdd{matteo.martinelli@inaf.it}
\emailAdd{rocco.dagostino@inaf.it}
\emailAdd{giulia.gianfagna@inaf.it}
\emailAdd{Carlos.Martins@astro.up.pt}

\abstract{Several assumptions at the foundation of the standard cosmological model have as a direct consequence a specific relation between cosmological distances, known as the distance duality relation, whose violation would be a smoking gun of deviations from standard cosmology. We explore the role of upcoming gravitational wave observations in investigating possible deviations from the distance duality relation, alongside the more commonly used supernovae. We find that, when combined with baryon acoustic oscillations, gravitational waves will provide similar constraining power to the combination of baryon acoustic oscillations and supernovae. Moreover, the combination of observables with different sensitivities to electromagnetic and gravitational physics provides a promising way to discriminate among different physical mechanisms that could lead to violations of the distance duality relation.
}

\newcolumntype{L}{>{\raggedright\arraybackslash}m{6cm}}  
\newcolumntype{C}{>{\centering\arraybackslash}m{3cm}}  
\newcolumntype{R}{>{\centering\arraybackslash}m{2cm}}

\begin{document}
\maketitle
\flushbottom

\section{Introduction}
\label{sec:intro}

The detection of gravitational wave (GW) signals 
has opened a new observational window into the Universe. Since the first detection by the LIGO-Virgo collaboration \cite{LIGOScientific:2016aoc}, GW have been used to investigate astrophysical phenomena \cite{Abbott_2023}, the nature of compact objects \cite{Abbott_properties} and the behaviour of General Relativity (GR) in extreme environments \cite{Abbott_2016}. The observation of GW170817 \cite{LIGOScientific:2017vwq}, originating from a binary neutron star merger with an associated electromagnetic counterpart, paved the way for the use of gravitational waves in the context of multimessenger cosmology. This allowed cosmologists to exploit GW observations as an independent probe for a measurement of the Hubble constant, $H_0$ \cite{LIGOScientific:2017adf}, opening a new avenue for cosmological investigation. Indeed, the combination of gravitational wave data with electromagnetic observations in GW170817 significantly improved the precision of the $H_0$ estimate — by a factor of 2–3 — compared to analyses based solely on gravitational waves \cite{Hotokezaka2018,Gianfagna_2024}.
In addition to this, the improved detection capabilities of future GW observatories, such as the Einstein Telescope  \cite{Abac:2025saz} or Cosmic Explorer \cite{Reitze:2019iox}, will open new frontiers in the study of the cosmos. 

A key application of GW in cosmology is their role as standard sirens, which allows one to directly measure the luminosity distance ($d_L$), without the need for a calibration, as it is instead necessary for electromagnetic standard candles such as type Ia supernovae (SNe).
The detection of GW signals and (when possible) of their electromagnetic counterparts enables independent constraints on cosmological parameters and, consequently, permits to investigate the nature of dark energy as well as fundamental questions about the gravitational interaction \cite{Cai:2016sby,Belgacem:2017ihm,DAgostino:2019hvh,Yang:2019vni,Lagos:2019kds,Bonilla:2019mbm,Mukherjee:2020mha,Mastrogiovanni:2020gua,Tasinato:2021wol,DAgostino:2022tdk,Antinozzi:2023yvl,Califano:2023aji}.
Besides allowing to obtain independent constraints on the standard cosmological model $\Lambda$CDM, which describes the Universe's history in terms of cold dark matter (CDM) and the cosmological constant ($\Lambda$) within the framework of GR, this new observational probe is also extremely promising to test the fundamental assumptions at the basis of the cosmological model.
Among the observational consequences of such fundamental assumptions, one that has gathered significant interest is the so-called Distance Duality Relation (DDR), or Etherington's relation \cite{Etherington}, i.e. the relation between $d_L$ and the angular diameter distance ($d_A$), which spawns directly from the assumptions of the standard cosmological model, and that is not violated in simple extensions.

Commonly, violations of the DDR have been explored in the electromagnetic (EM) sector, where they could signal the necessity of new physics, such as the non-conservation of the photon number, deviations from a metric theory of gravity, or unusual photon propagation effects \cite{Avgoustidis2009}. 
Several works have investigated how current and upcoming measurements of luminosity and angular distances can be combined together to obtain tight constraints on DDR deviations \cite{EUCLID:2020syl,Favale:2024sdq,Keil:2025ysb,Alfano:2025gie}, while others have focused on the search of stand-alone probes for such a non-standard behaviour, such as strongly lensed systems \cite{Renzi:2020bvl,Tang:2024zkc}. Furthermore, violations of the DDR have been explored also in the context of the $H_0$ tension \cite{Perivolaropoulos:2021jda,Abdalla:2022yfr,DiValentino:2025sru}, investigating how the latter could be an observational consequence of a departure from the DDR
\cite{Renzi:2021xii,Teixeira:2025czm}. Violations of the DDR have also been investigated in the context of addressing potential statistical correlations among cosmographic parameters \cite{Jesus_2025}.

Even though most of the investigations done in the context of DDR  have focused on EM probes, the expected improvement in GW measurements has prompted cosmologists to enquire how these can be used to further test such a fundamental relation \cite{yang2019constraintscosmicdistanceduality}. In particular, recent results have highlighted how GW could be used to break the degeneracy between violations of the DDR and the impact of modified gravity theories on observations \cite{Hogg:2020ktc}, while it has also been shown that combining measurements of EM and GW distances could lead to tight and model-independent tests of the DDR \cite{Matos:2023jkn}. Finally, other studies have shown that GW events from current and upcoming detectors can serve as a model-independent test of General Relativity by probing the relationship between the luminosity distance in the gravitational and EM sectors
\cite{afroz2024prospectprecisioncosmologytesting, afroz2024modelindependentprecisiontestgeneral, afroz2024modelindependentprecisiontestgeneral2}.

In this work, we focus on assessing the improvement in constraining power on DDR violations when GW, in the form of bright sirens, are used alongside EM probes to bound parameterised deviations from standard behaviour. We show, however, how this implies the assumption that the mechanism responsible for possible violations affects the gravitational and luminosity distances in the same way. As this is not necessarily the case, with models breaking the DDR differently in the EM and GW sectors \cite{Avgoustidis2010}, a key result of this work is to show the role of GW in distinguishing between these possibilities, offering a way to identify whether a hypothetical violation of the DDR arises from electromagnetic or gravitational origins. Given the sensitivity required for this kind of investigation, we restrict our analysis to upcoming surveys in both the EM and GW sectors. In particular, we rely on the expected sensitivity of the Einstein Telescope \cite{abac2025scienceeinsteintelescope} for the measurement of GW distances, of SKAO for estimates of the angular diameter distance through baryon acoustic oscillations (BAO) measurements \cite{SKAO_gen}, and on the expected SN observations of the Vera C. Rubin observatory \cite{Bianco2022}, which will give information on the EM luminosity distance.

This paper is organised as follows. In \autoref{sec:theory} we outline the theoretical assumptions that are connected with DDR, also discussing possible models that could lead to violations of such a relation, and the parameterised approaches commonly used to constrain any departure from it. In \autoref{sec:data} we describe the methodology used to obtain synthetic datasets of the future surveys that we are interested in, while \autoref{sec:method} contains a description of the approach used to compare such datasets with theoretical predictions, allowing us to obtain the constraints we then present in \autoref{sec:results}. Finally, we draw our conclusions in \autoref{sec:conclusions}.

\section{Distance duality relation and possible violations}\label{sec:theory}

In any cosmological model based on a metric theory of gravity, the Etherington relation, or distance duality relation, implies that distance measures are unique \cite{Etherington}. Specifically, at any redshift $z$, a well-defined relation connects the angular diameter distance $d_A(z)$, which characterizes the apparent size of a distant source, to the luminosity distance $d_L(z)$, which is determined by the flux received from that source.
In the framework of GR, the number of particles emitted from the source is conserved, and they travel on null geodesics (i.e., they are massless). In this case, the DDR implies
\begin{equation}\label{eq:std_DDR}
    d_{\rm L}(z) = (1+z)^2d_{\rm A}(z)\, .
\end{equation}
This relation is crucial for cosmology, as it allows us to tie together two quantities that can be measured and can be directly connected with the underlying cosmological model. Indeed, in an expanding Universe described by a spatially flat Friedmann-Lema\^itre-Robertson-Walker metric
\begin{equation}
    {\rm d}s^2 = -c\, dt^2+a^2(t)dx^2\,,
\end{equation}
where $a=1/(1+z)$ is the scale factor, one can define the comoving distance $d_C(z)$ as 
\begin{equation}
    d_{\rm C}(z)= \int_0^z \frac{c\,dz'}{H(z')}\,,
\end{equation}
where $H\equiv \dot{a}/a$ is the Hubble parameter.
From this, one can obtain the angular diameter distance as
\begin{equation}
    d_{\rm A}(z)=\frac{1}{(1+z)}d_{\rm C}(z)\,.
\end{equation}

If the DDR holds, one can easily obtain the luminosity distance from Eq.~\eqref{eq:std_DDR}. However, when one of the assumptions outlined above is violated, it will lead to departures from the standard Etherington relation, which can be encoded in a function $\eta(z)$ defined as
\begin{equation}
    \eta(z)\equiv\frac{d_{\rm L}(z)}{(1+z)^2d_{\rm A}(z)}\,,
    \label{eq:etapar}
\end{equation}
which is constant and equal to unity under the DDR assumption. Moreover, regardless of any possible violations at high redshifts, $\eta(z)$ must approach unity in the limit $z \rightarrow 0$, since any object at zero distance should appear at zero distance, regardless of how the distance is measured.

%\subsection{Violation mechanisms and impact on observables}\label{sec:obs}

While DDR violations could be associated with the failure of multiple assumptions at the basis of standard cosmology, the violation of different assumptions would lead to distinct observational signatures. 
In particular, it is important to stress that the luminosity distance can be connected to electromagnetic observables (e.g., type Ia SNe), but also to measurements of GW, which would not rely on the observations of photons. For such a reason, we distinguish here between two types of luminosity distances: electromagnetic $d_{\rm L}^{\rm EM}(z)$ and gravitational $d_{\rm L}^{\rm GW}(z)$.
While these two quantities coincide in standard cosmology, a violation of DDR would in general affect the two distances differently, depending on the mechanism leading to the breakdown of cosmological assumptions.
As an example, in models where photons can decay into another particle that remains unobserved, there is an effective violation of photon number conservation, which leads to a departure of $\eta^{\rm EM}(z)$ from unity \cite{Avgoustidis2010}. At the same time, the propagation of GW is not affected by such a mechanism and, therefore, $\eta^{\rm GW}=1$.
On the other hand, mechanisms that leads to violation of the DDR through modifications of the space-time geometry, e.g. in torsion extensions of GR \cite{Bolejko:2020nbw,DAgostino:2025kme}, can impact the evolution of both EM and GW distances, thus leading to a different signature in the GW sector with respect to mechanisms that only affect photons.

More generally, several classes of modified gravity theories predict a deviation of the GW luminosity distance from the standard form. For instance, scalar-tensor theories (including Horndeski and beyond-Horndeski models) can alter GW propagation through a time-varying Planck mass or additional friction terms, leading to a redshift-dependent modification of $d_L^{\rm GW}(z)$ \cite{Ezquiaga:2017ekz,Belgacem:2018lbp,Nishizawa:2019rra}. Similarly, in massive gravity and bimetric theories, a modified dispersion relation for GWs or coupling to an auxiliary metric can lead to observable effects in the GW sector \cite{deRham:2010kj,Hassan:2011zd,Gumrukcuoglu:2012wt,deRham:2014zqa}. These theoretical frameworks provide well-motivated mechanisms in which the breaking of the DDR can be tested with multi-messenger observations. We therefore emphasize that the comparison between $d_L^{\rm{EM}}(z)$ and $d_L^{\rm{GW}}(z)$ offers a powerful diagnostic tool to discriminate among the physical origins of a possible DDR violation, distinguishing, e.g., between photon-specific effects and modifications to the gravitational sector.

\subsection{Phenomenological deviations from the DDR}

To take into account all the different mechanisms that could lead to a violation of the DDR, it is possible to adopt a phenomenological description of this relation. Commonly, this can be described as
\begin{equation}\label{eq:epspar}
    d_{\rm L}(z) = (1+z) ^{2+\epsilon(z)}d_{\rm A}(z)\,,
\end{equation}
and, consequently, Eq.~\eqref{eq:etapar} can be rewritten as
\begin{equation}
    \eta(z) = (1+z)^{\epsilon(z)}\,.
    \label{eq:eta0}
\end{equation}
The deviation from the standard DDR is encoded in $\epsilon(z)$, which is commonly taken as a constant. The constraints obtained with current data are of the order of the order of $10^{-1}$ in the $2\sigma$ range \cite{EUCLID:2020syl}. 

Given their impact on a fundamental quantity such as $d_{\rm L}(z)$, potential violations of the DDR can, in principle, be tested and constrained using observational data. If the reconstructed functions from Eq.~\eqref{eq:etapar} and Eq.~\eqref{eq:epspar} deviate from their standard values $\eta(z) = 1$, $\epsilon(z) = 0$, this might indicate a breakdown of the standard cosmological model, implying that one or more of its fundamental assumptions may no longer hold.

In order to obtain constraints on these functions, a common approach is to parameterize them. In the easiest approach, the function $\epsilon(z)$ is assumed to be a constant, with its value $\epsilon_0$ being a free parameter to be constrained with observational data \cite{Avgoustidis2009,Avgoustidis2010,EUCLID:2020syl}. However, with the increase in precision of the available data, different approaches have been used, trying to also detect possible redshift trends for $\epsilon(z)$ or $\eta(z)$, e.g. by binning in redshift \cite{EUCLID:2020syl} or applying machine learning reconstruction methods such as Gaussian Processes or Genetic Algorithms \cite{Hogg:2020ktc,Renzi:2020bvl}.
While this phenomenological definition applies in general, it is important to explore the theoretical models leading to violations of the DDR, as their impact on cosmological observables will be different.

\subsubsection{Padé parametrization}

It has been argued that the trend of DDR violations produced by physical models should be monotonic in redshift \cite{Avgoustidis:2009ai}. While a monotonic trend can be reproduced by the parametrization we presented above, such a behaviour can lead to issues when modelling theoretical predictions for high redshift observables, which would require studying physical mechanisms in such a modified scenario. It is therefore common to work with parametrizations that include a cut-off of the DDR violation, neglecting the impact of DDR violations at high redshift. In order to avoid an artificially sharp cut-off, it is possible to rely on approaches that naturally lead the modifications introduced to vanish at high redshift.

A robust approach leading to such an effect 
involves the use of rational polynomials \cite{Wei:2013jya,Capozziello:2017nbu}. A notable example in this respect is offered by Padé approximants, which have been applied to cosmographic reconstructions of the Hubble expansion rate to heal the convergence issues inherent in Taylor series expansions of the luminosity distance \cite{Capozziello:2022jbw,Capozziello:2020ctn}, as they offer both stability at large cosmological distances and improved convergence properties \cite{Aviles:2014rma,DAgostino:2023cgx}.

The Padé approximant method has found broad applications across various contexts, ranging from kinematic analyses in standard cosmology to investigations within modified gravity frameworks \cite{Capozziello:2017ddd,Capozziello:2019cav,Capozziello:2022wgl,Capozziello:2023ccw}. 
Specifically, one can construct the $(n,m)$ Padé approximant of a function $f(z)$ as 
\begin{equation}
P_{m,n}(z) = \dfrac{\displaystyle\sum_{j=0}^n b_j z^j}{\displaystyle\sum_{i=0}^m a_i z^i}\,,
\end{equation}
where the orders of polynomials in the numerator ($n$) and in the denominator ($m$) should be suitably chosen, depending on the specific problem under study. The unknown coefficients $a_i$ and $b_j$ are determined by solving the following system: 
\begin{equation}
\left\{
\begin{aligned}
&a_i=\sum_{k=0}^i b_{i-k}\ c_{k} \ ,  \\
&\sum_{j=1}^m b_j\ c_{n+k+j}=-b_0\ c_{n+k}\ , \quad k=1,\hdots, m 
\end{aligned}
\right .
\label{eq:Pade_system}
\end{equation}
where the coefficients $c_k$ are the expansion coefficients of the Taylor series of the starting function, $f(z)=\sum_{k=0}^\infty c_k z^k$. 

Motivated by the optimal convergence properties and statistical performances of the (2,1) Padé polynomial as demonstrated in \cite{Capozziello:2020ctn}, we consider the following parametrization:
\begin{equation}
    \eta_{(2,1)}=\frac{1+a_1z+a_2z^2}{1+b_1z}\,.
    \label{eq:eta_new}
\end{equation}
We can constrain the coefficient $b_1$ by requiring that $\eta(z)\to 1$ at $z\to z_*$, where $z_*$ is the redshift of the last scattering surface. This condition is to guarantee the matching with the standard cosmological prediction at very high redshifts. In doing so, we find
\begin{equation}
    b_1=a_1+a_2z_*\,.
    \label{eq:b1_new}
\end{equation}
Additionally, we can obtain the coefficients $a_1$ and $a_2$ by equating the Taylor series around $z=0$ of Eq.~\eqref{eq:eta0} with that of Eq.~\eqref{eq:eta_new}. We have, respectively,
\begin{align}
    &\eta\simeq 1+z\,\epsilon\, +\frac{z^2}{2}  (\epsilon -1) \epsilon\,, \\
    &\eta_{(2,1)}\simeq 1 + z(a_1 - b_1) + z^2(a_2 - a_1 b_1 + b_1^2)\,.
\end{align}
Thus, comparing the same powers of $z$ and taking into account Eq.~\eqref{eq:b1_new}, we obtain
\begin{equation}
    a_1=\frac{1+\epsilon}{2}-\frac{1}{z_*}\,, \quad a_2=-\frac{\epsilon}{z_*}\,, \quad b_1 =\frac{1-\epsilon}{2}-\frac{1}{z_*}\,.
\end{equation}
Inserting the above expressions into Eq.~\eqref{eq:eta_new} provides us with
\begin{equation}
    \eta_{(2,1)}=\frac{z \left[2(1+\epsilon\, z) -z_* (1+\epsilon)\right]-2 z_*}{z\left[2-z_*(1-\epsilon)\right]-2z_*}\,,
\end{equation}
which can be expanded at first order around $\epsilon=0$, to finally obtain 
\begin{equation}
    \eta_{(2,1)}\simeq 1+\frac{2z(z-z_*)\epsilon}{2(z-z_*)-z\, z_*}\,.
    \label{eq:eta_new_series}
\end{equation}
Note that Eq.~\eqref{eq:eta_new_series} is ill-defined for $z=\frac{2z_*}{2-z_*}$. However, since $z_*\approx 1100$, this undesirable behaviour occurs for $z<0$, which corresponds to the future and lies outside the redshift range considered in the present study. 

\begin{figure}[h!]
\centering        \includegraphics[width=1\textwidth]{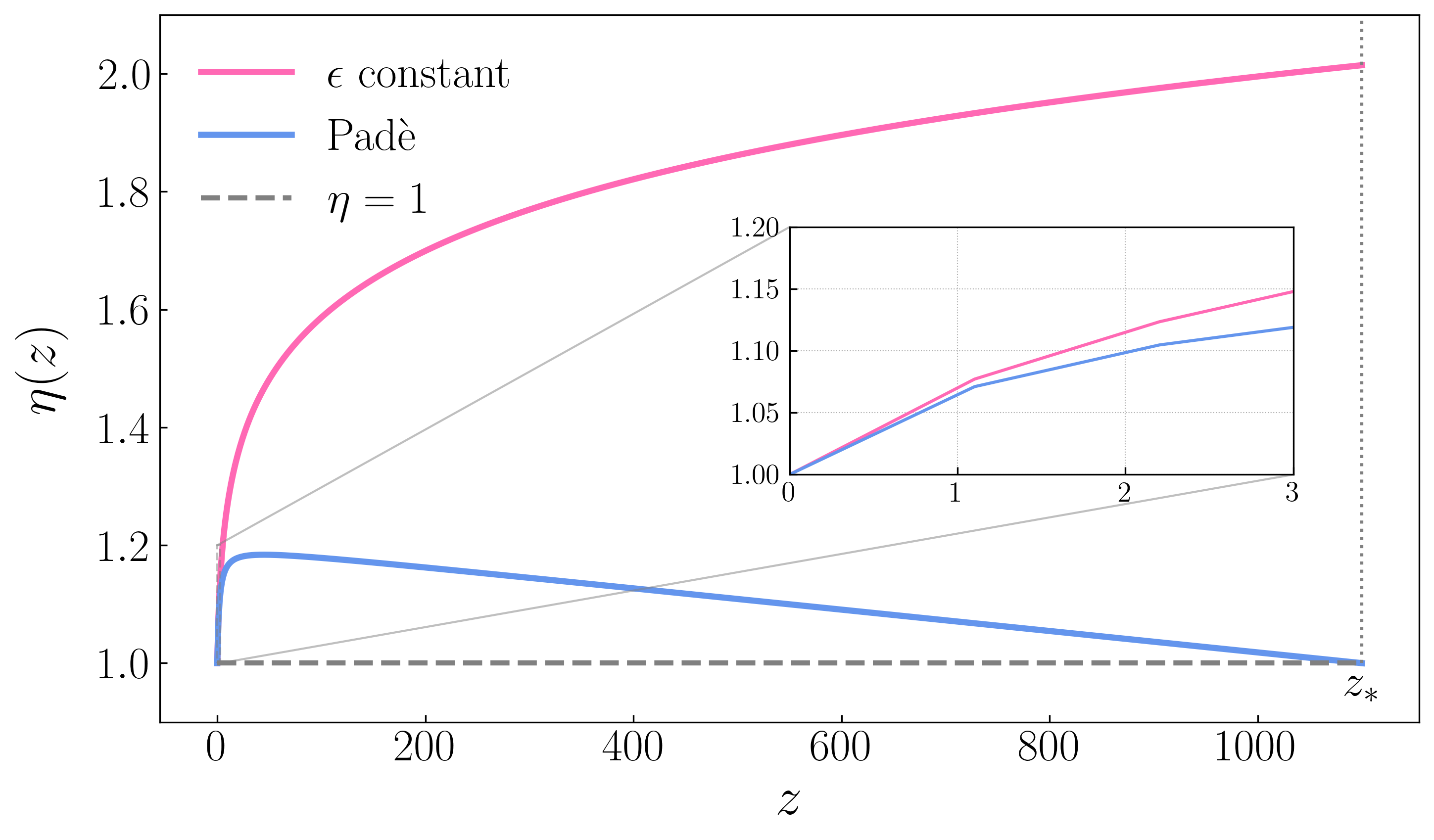}
\caption{ Comparison of the behaviour of $\eta(z)$ using the Padé expression (blue) and the constant $\epsilon$ expression. The inset shows a zoomed-in view of the redshift range relevant for this work. The horizontal grey line marks $\eta = 1$, and the vertical dashed grey line indicates the redshift of the Last Scattering Surface $z_*$.}
\label{fig:pade_eta}
\end{figure}

In \autoref{fig:pade_eta}, we show the behavior of $\eta(z)$ when modeled using a Padé approximation compared to a simpler expression given in Eq.~\eqref{eq:eta0}. The inset in the figure highlights the differences between the two approaches within the redshift range relevant to our analysis.

\section{Cosmological data}\label{sec:data}

The purpose of this work is to understand the possible constraints on violations of the DDR that will be achievable with future surveys. In particular, we are interested in the possibility of distinguishing between different theoretical models when gravitational waves are used alongside electromagnetic data.

We therefore use two different observables to obtain measurements related to luminosity distances. On the one hand, gravitational waves observations will provide measurement of $d_{\rm L}^{\rm GW}(z)$, i.e. the distance that can be measured from the merging of compact binaries, while SN-Ia observations are related to $d_{\rm L}^{\rm EM}(z)$, through their use as standardisable candles. 
While these two observables are in principle enough to obtain constraints on the violation of DDR for the two sectors, the DDR parameters are very degenerate with other standard cosmological parameters, e.g. with the energy density of matter $\Omega_{\rm m}$ \cite{EUCLID:2020syl}. For such a reason, in order to obtain tight constraints, we include in our set of observables also BAO observations. The latter provide measurements of quantities that are related to $d_A(z)$ and $H(z)$, which are not affected by violations of the DDR. For this reason, BAO can act as an anchor for the constraints obtained with GW and SN, breaking the degeneracies with cosmological parameters.
In this work, we utilize these three %different 
observables, with a particular emphasis on future datasets. In \autoref{sec:bao}, \autoref{sec:sn}, \autoref{sec:gw} we enter into the details of each observable, in particular on how distance measurements can be extracted from observations, and we explain the methodology that we follow to simulate data from upcoming surveys.

\subsection{Baryon acoustic oscillations} \label{sec:bao}

BAO measurements provide information on a set of distances (the transverse distance $D_M$, the Hubble distance $D_H$ and the angle-average distance $D_V$) rescaled by the sound horizon at the drag epoch $r_d$, which can be obtained as
\begin{equation}\label{eq:sound_horizon}
    r_d=\int_{z_d}^\infty {{\rm d}z\frac{c_s(z)}{H(z)}}\,,
\end{equation}
where $c_s(z)$ is the speed of sound. This, prior to recombination, can be expressed as
\begin{equation}\label{eq:sound_speed}
    c_s(z)=\frac{c}{\sqrt{3\left(1+\frac{3\rho_{\rm b}(z)}{4\rho_\gamma(z)}\right)}}\,,
\end{equation}
where $\rho_{\rm b}$ and $\rho_\gamma$ are the energy densities for baryons and relativistic particles, respectively.

\begin{figure*}[h!]
    \centering
    \begin{subfigure}[b]{0.48\textwidth}
        \centering
        \includegraphics[width=\textwidth]{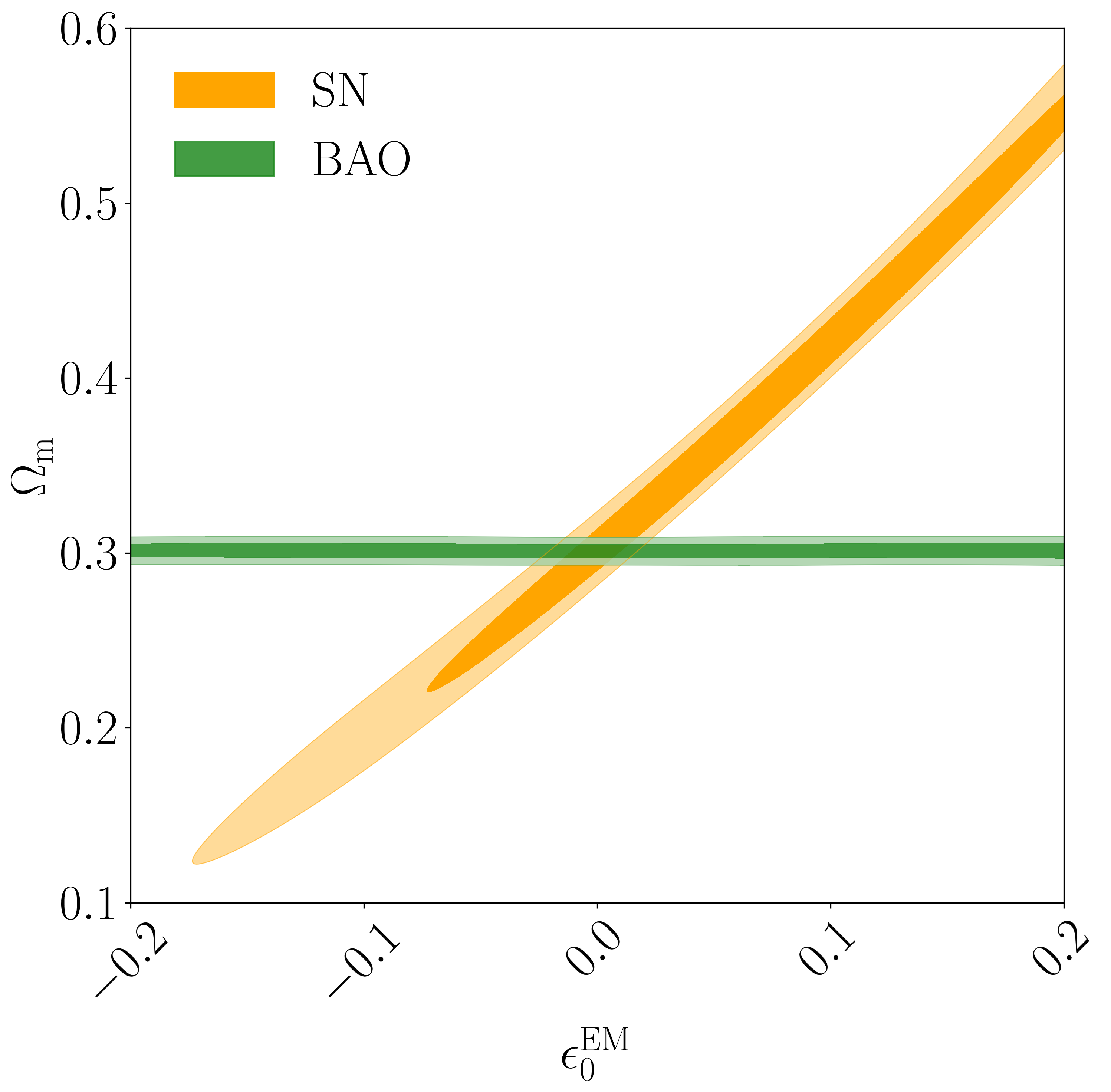}
        \caption{}
        \label{fig:BAO+SN}
    \end{subfigure}
    \hfill
    \begin{subfigure}[b]{0.48\textwidth}
        \centering
        \includegraphics[width=\textwidth]{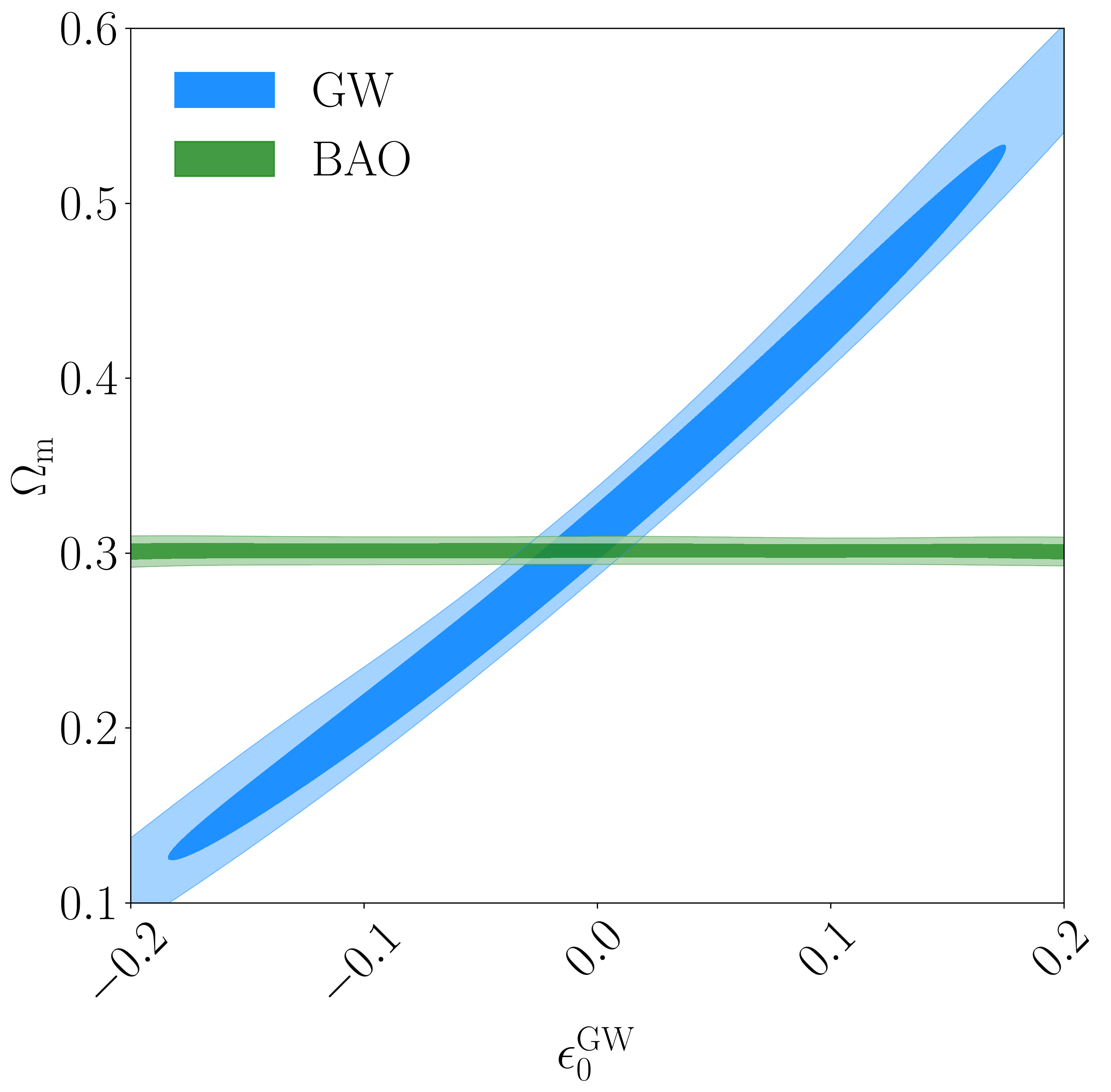}
        \caption{}
        \label{fig:BAO+GW}
    \end{subfigure}
\caption{\textbf{(a) :} Two-dimensional  contours on $\Omega_m$ and $\epsilon_0^{EM}$ using BAO (green) and SN (orange). \textbf{(b) :} Two-dimensional  contours on $\Omega_m$ and $\epsilon_0^{GW}$ using BAO (green) and GW (blue). Both results are obtained using the synthetic data described in \autoref{sec:data} and the methodology described in \autoref{sec:method}.}
    \label{fig:BAO_deg_SNGW}
\end{figure*}

BAO observations measure the quantities $D_M/r_d$, $D_H/r_d$, $D_V/r_d$, where the distances in a flat Universe are defined as \cite{DESI:2024mwx}:
\begin{equation}\label{eq:bao_distances}
D_M(z) = \frac{c}{H_0}\int_0^z \frac{{\rm d}z'}{E(z')}, \quad
D_H(z) = \frac{c}{H(z)}, \quad
D_V(z) = \left[zD_M^2(z)D_H(z)\right]^{1/3}.
\end{equation}
As said before, the BAO measurements are not sensitive to violation in the DDR relation but they help in breaking the degeneracy between the DDR parameter $\epsilon$ and $\Omega_m$, as shown in previous works \cite{EUCLID:2020syl} and in \autoref{fig:BAO_deg_SNGW}

In this work, we explore the constraining power of future BAO measurements, specifically those expected from the SKA Observatory (SKAO) \cite{SKAO}. In particular, we simulate the BAO measurements assuming the optimistic configuration, labelled as SKA2 in \cite{SKAO_gen},
where the 21-cm spectral line will be probed to extract BAO information in the redshift range $z\in\left[0.2,2\right]$, divided in $18$ bins of width $\Delta z=0.1$.

It must be pointed out that BAO measurements do not provide information directly on the distances of Eq.~\eqref{eq:bao_distances}, but rather on their ratio with the sound horizon $r_d$. As it can be seen from Eq.~\eqref{eq:sound_horizon} and Eq.~\eqref{eq:sound_speed}, disentangling such a degeneracy would require information on the baryonic energy density $\rho_b$, which cannot be provided by BAO measurements. Indeed, the use of external priors, such as those coming from Big Bang Nucleosynthesis (BBN) \cite{BBN}, allows to probe directly the BAO distances, but we choose here to avoid the use of external prior to be consistent with the choice of neglecting the SH0ES calibration for the SNe, as we will discuss in \autoref{sec:sn}. While an early-time DDR violation would impact both the BBN process and the BAO scale, we neglect these effects here as they lie outside the scope of this work.
%as it would require us to model the impact of a DDR violation on early Universe physics.
For such a reason, we simulate a dataset for $D_M/r_d$ and $D_H/r_d$, propagating to these quantities the relative errors on $d_A$ and $H$ presented in Figure 4 of \cite{Bull:2015lja}.

\subsection{Supernovae}\label{sec:sn}
To constrain the luminosity distance from an electromagnetic perspective, $d_L^{\rm EM}(z)$, we use SNe, which provide a direct measurement of the apparent magnitude, $m(z)$. This is related to the luminosity distance through the well-known relation:
\begin{equation}
   m(z) = M_B + 5 \log_{10} \left(\frac{d_L^{\rm EM}(z)}{\text{Mpc}}\right) + 25,
   \label{eq:mz}
\end{equation}
where $M_B$ is the SN absolute magnitude. The luminosity distance $d_L^{\rm EM}(z)$ encodes cosmological information, while $M_B$ can be considered as a calibration parameter. 
A key aspect is that $M_B$ is generally degenerate with the Hubble constant $H_0$, unless an external information is added. We will describe the strategy adopted to overcome this issue in \autoref{sec:method}.
\newline
For our analysis, we simulate future SN data from the Legacy Survey of Space and Time\footnote{\href{https://rubinobservatory.org/for-scientists/rubin-101/key-numbers}{https://rubinobservatory.org/for-scientists/rubin-101/key-numbers}} (LSST) \cite{Bianco2022}. We simulate $N_{\rm SN}=8800$ events, that corresponds to LSST expected number of events observed in the redshift range $z=0.1-1.0$, each with an error given by \cite{2010ApJ...709.1420G,Astier:2014swa}
\begin{equation}
    \sigma_i^2=\delta\mu_i^2+\sigma^2_{\rm flux}+\sigma^2_{\rm scat}+\sigma^2_{\rm intr},
    \label{eq:err_SN}
\end{equation}
where $\sigma_{\rm flux}=0.01$ is the contribution to the observational error due to the flux, $\sigma_{\rm scat}=0.025$ to the scattering and $\sigma_{\rm intr}=0.12$ are the intrinsic uncertainties.

\subsection{Standard Sirens}\label{sec:gw}
The last observable that we include are GW, and in particular the Standard Sirens \cite{Holz_2005, Nissanke2010}. GW represent a perturbation of spacetime that, in GR, propagates according to the equation 
\begin{equation}
    h_{+,\times}''(\tau,k) +2aHh_{+,\times}'(\tau,k)+k^2h_{+,\times}(\tau,k)=0,
\end{equation}  
where $h_{+,\times}(\tau,k)$ is the strain amplitude for the two polarization modes $+,\times$, and $\tau$ is the conformal time. It can be demonstrated \cite{1986Natur.323..310S} that the strain amplitude $h_{+,\times}$ is inversely proportional to  $d_L^{\rm GW}(z)$, meaning that the GW strain decreases with distance from the source as  
$$ h \propto \frac{1}{d_L^{\rm GW}(z)}. $$  
By accurately modelling the GW waveform, which depends on the masses, the orientation of the merging system, the spins of the objects and the distance, one can infer $d_L^{\rm GW}$ directly from the observed amplitude without relying on the traditional cosmic distance ladder. If an EM counterpart or an identified host galaxy also gives information about the redshift $z$, combining $d_L^{\rm GW}$ and $z$ provides a direct measurement of $H(z)$. These are known as Standard Sirens \cite{Nissanke2010}.

We simulate standard siren events by focusing on the merger of binary Neutron Stars (BNS). Our simulations are based on the expectations for the Einstein Telescope (ET) \cite{abac2025scienceeinsteintelescope}, a forthcoming third-generation ground-based GW detector \cite{Branchesi_2023}. We assume a triangular configuration with 10 km arms, located in Sardinia.
We generate $\rm N=20000$ GW events, also known as injected signals, from BNS mergers with a rate modelled as \cite{Cutler:2009qv,Hogg:2020ktc}
\begin{equation}\label{eq:merger_rate}
    R(z) = 
    \begin{cases} 
        1 + 2z & \text{if } z \leq 1, \\
        \frac{3}{4}(5-z) & \text{if } 1 < z < 5, \\
        0 & \text{if } z \geq 5.
    \end{cases}
\end{equation}
For the position in the sky we assume that $\rm sin \theta_{\nu}$, where $\theta_\nu$ is the inclination angle, is uniformly distributed in the range $[-1,1]$; while for the redshift of the events we use a distribution

\begin{equation}
    P(z) \propto  \frac{ R(z)4 \pi d_C(z)^2}{H(z)(1+z)}.
\end{equation}

The events are simulated exploiting a modified version of the \texttt{darksirens}\footnote{\href{https://gitlab.com/matmartinelli/darksirens}{https://gitlab.com/matmartinelli/darksirens}} code \cite{darksirens}, which implements the merger rate of Eq.~\eqref{eq:merger_rate} and a monochromatic mass for the neutron stars of $1.4\ M_\odot$.

We compute the luminosity distance of a GW event using Eq.~\eqref{eq:epspar} and Eq.~\eqref{eq:eta0}.
The uncertainties on the luminosity distance $d_L^{\rm GW}$ and the detectability of the signals are evaluated using the public \texttt{GWFish}\footnote{\href{https://github.com/janosch314/GWFish}{https://github.com/janosch314/GWFish}} code \cite{Dupletsa_2023}.
For what concerns the detectability, we first perform a cut in the data based on the SNR ratio, considering an event to be detected if its signal-to-noise ratio (SNR) is larger than eight ($\rm SNR > 8$), and then we analyze the EM counterpart detectability.
The number of detected BNS mergers increases if ET is in a network of third-generation GW detectors. As an example, ET plus one Cosmic Explorer \cite{Reitze:2019iox} would increase of a factor $\sim3$ the number of detected BNS, and a factor $\sim4$ with two Cosmic Explorers. We also note that even changing the ET configuration (two L-shaped detectors instead of a triangle) can lead to an almost 3 times higher detection rate \cite{Branchesi_2023}. Given all these uncertainties, we choose to focus only ET, with a triangular configuration.

For the EM counterparts, we expect mainly two types of emission: a Gamma Ray Burst (GRB) and a Kilonova \cite{metzger2017welcomemultimessengereralessons}. Following the merger, a highly collimated and relativistic jet is formed, which emits the GRB prompt emission and its afterglow. The prompt emission is detected right after the merger, at high (gamma-ray) energy, and with a duration below 2 sec. Once the jet reaches the interstellar medium, it starts to decelerate and spread, producing the afterglow. The latter is mainly synchrotron emission spanning the whole spectrum from X-rays to radio wavelengths and can last for even years after the merger. 

A small amount of fast-moving neutron-rich ejecta emits an isotropic thermal emission, the kilonova, peaking in the infrared. A larger mass of neutron-free wind along the polar axis produces the kilonova emission peaking at optical wavelengths. 

The detectability of these EM sources will depend on the luminosity, the redshift and the inclination angle. 

In this work, we focus on the afterglow emission and we assume that each GW event produces a GRB (even if this is not necessarily true \cite{Metzger_counterpart}).
%a given). 
In particular, we use the luminosity distance and the inclination from the GW counterparts to generate the afterglow flux using the code \texttt{afterglowpy}\footnote{\href{https://github.com/geoffryan/afterglowpy}{https://github.com/geoffryan/afterglowpy}} \cite{Ryan_2020}. The flux also depends on the energetics and the mycrophysics of the event, which we fix to the one estimated in population studies of short GRBs \cite{RoucoEscorial_2023, Fong_2015}. In particular, we assume a Gaussian structured jet and we fix the jet opening angle to 6${\rm ^o}$, the isotropic-equivalent kinetic energy $E_0$ to $4\times10^{51}$ erg, the circum-burst density $n_0$ to $6.4\times10^{-4}$ cm$^{-3}$, the fraction of energy in the magnetic field $\epsilon_B$ and in the electrons $\epsilon_e$ both to 0.1, and the slope of the electron distribution $p$ to 2.3. 

\begin{figure}[h!]
    \centering
    \includegraphics[width=1.\linewidth]{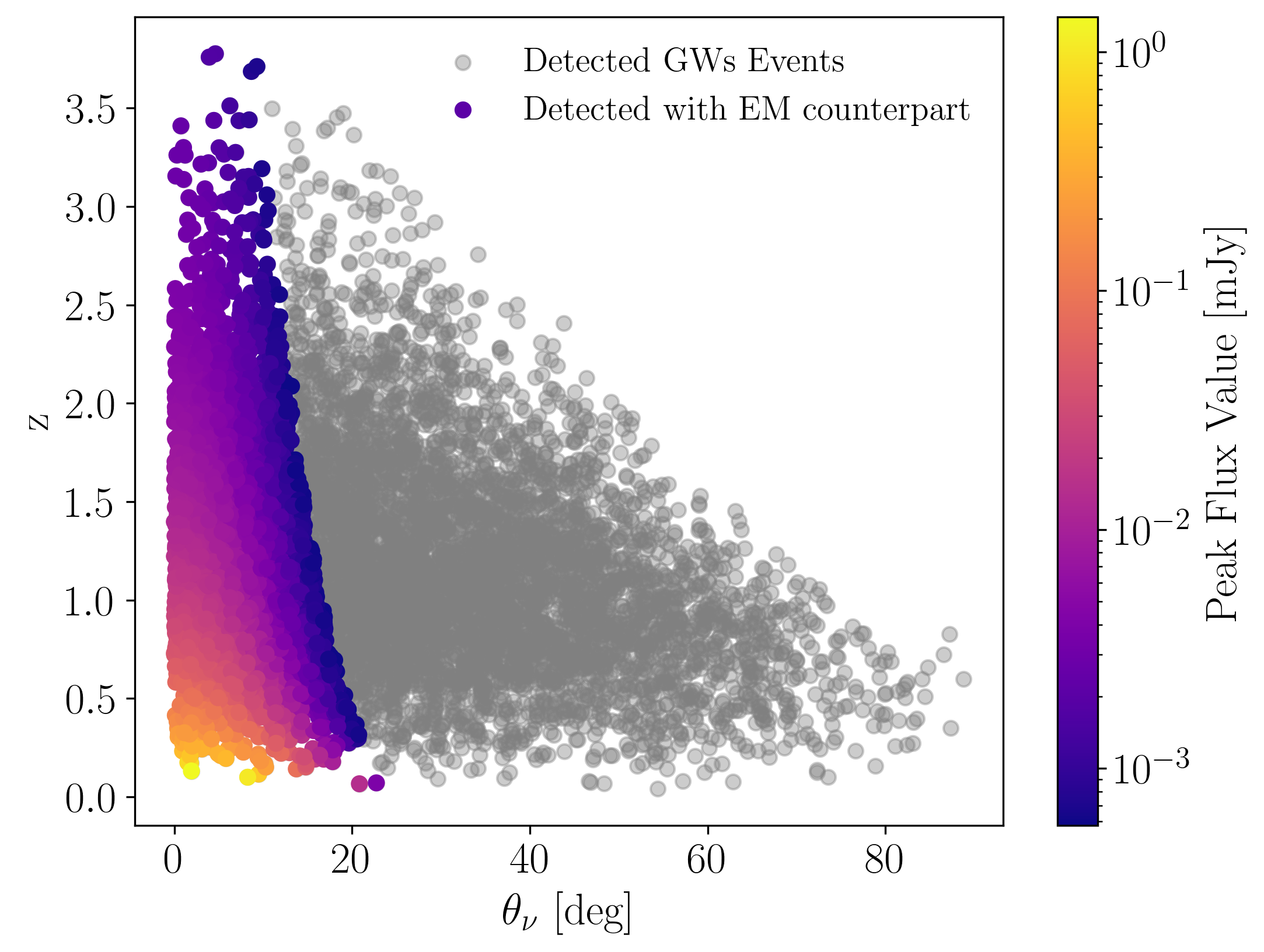}
    \caption{Detected GW events as a function of redshift and inclination angle $\theta_\nu$, along with their corresponding peak flux. The events are simulated assuming a fiducial cosmology from Pantheon+SH0ES \cite{Riess_2022}, with $H_0=73.4$ and $\Omega_m=0.306$. The grey points represent events detected by ET that survive the SNR cut, while the coloured points indicate events that would also be detected with an associated electromagnetic counterpart.}
    \label{fig:Detected_GW}
\end{figure}

As a reference, we use the sensitivity of the Vera Rubin Observatory (LSST), which will become operational at the end of 2025. It will reach a sensitivity of $24.5$ magnitudes in the R band \cite{Bianco2022}, so a flux threshold of about $F_{\rm th}=5.8\times10^{-4}\,\, \rm mJy$. Taking into consideration all these characteristics we obtain that around $40 \%$ of the GW events detected would have an EM counterpart, this can be seen in \autoref{fig:Detected_GW}. As expected, the smaller the inclination, the higher the chance to detect the source.

We note that we choose to generate the afterglow fluxes in the optical band so that also the redshift measurement is most likely feasible. Moreover, this is a conservative approach, as we do not take into account the kilonova emission, which should be easier to detect even at high inclinations, due to its wider opening angle. At the same time, the kilonova is expected to be dimmer than the afterglow, so it would be visible up to lower redshifts. In addition, we do not take into account the distribution in the sky of the events, in fact LSST will be able to observe half of the sky.

\section{Methodology}\label{sec:method}
The goal of this work is to assess the potential of incorporating GW data, which provide independent measurements of the luminosity distance $d_L$, to test the DDR described in Eq.~\eqref{eq:eta0}. In this section, we describe the methodology applied and the different scenarios that have been tested.
\subsection{Reconstructing the posterior distribution}
In order to constrain possible deviations from the standard DDR, we use statistical methods to obtain constraints on the $\eta(z)$ function, which we parameterize following the discussion of \autoref{sec:theory}.
This process is done by reconstructing the posterior distribution of $\epsilon$, together with other cosmological parameters that will be discussed in \autoref{sec:models}. This reconstruction is grounded in Bayesian statistics \cite{bayesian}, with the comparison between data and theoretical predictions handled by the public likelihood code
\texttt{Cobaya}\footnote{\href{https://github.com/CobayaSampler/cobaya}{https://github.com/CobayaSampler/cobaya}} \cite{Cobaya}, for which we construct external likelihood modules able to handle the datasets discussed in \autoref{sec:data}.

For what concerns the theoretical modelling, 
we obtain the standard cosmological quantities through \texttt{CAMB}\footnote{\href{https://camb.readthedocs.io/}{https://camb.readthedocs.io/}} \cite{Lewis:1999bs,2012JCAP...04..027H}, 
interfaced with a theory code that includes the expression for $\eta$ and the modifications to $d_L$.
The setup used in this work, made publicly available as \texttt{CANDI}\footnote{\href{https://github.com/chiaradeleo1/CANDI}{https://github.com/chiaradeleo1/CANDI}} \cite{Lewis:1999bs,2012JCAP...04..027H}, allows \texttt{Cobaya} to perform Markov Chain Monte Carlo (MCMC) sampling to reconstruct the posterior distribution \cite{Lewis:2002ah,Lewis:2013hha}.

\subsection{Models and sampled parameters} \label{sec:models}
As explained in \autoref{sec:data}, we simulate next-generation datasets for BAO, SN and Standard Sirens.
By analyzing different combinations of these simulated datasets in different scenarios, we aim to evaluate the sensitivity of future observations to various mechanisms that could lead to a breakdown of the DDR. In particular, we focus on three different combinations:
\begin{itemize}
    \item Both SN and GW are consistent with the DDR and no violation happens ($\Lambda$CDM Universe);
    \item Both SN and GW violate the DDR, and this could be addressed by some modifications in the theory of gravity (full-breaking Universe);
    \item Only SN violate the DDR, and this would be the case of dust or decay of photons into non-standard model particles (EM-breaking Universe).
\end{itemize}

We generate simulated observations corresponding to these scenarios, using the fiducial values from Pantheon+SH0ES \cite{Riess_2022} shown in \autoref{tab:fiducials}.
We analyze these simulated observations sampling the DDR violations parameters, together with the cosmological ones, assuming the prior distributions reported in \autoref{tab:prior}.
\begin{table*}[!ht]
\caption{Fiducial values of the cosmological parameters used for the generation of the simulated datasets in the standard and breaking scenarios. Common parameters are listed separately, while parameters specific to each observable are provided below the corresponding observable.
}
\centering
\begin{tabularx}{\textwidth}{m{5.5cm}m{2.5cm}m{7cm}}
\toprule
\textbf{Description} & \textbf{Symbol} & \textbf{Fiducial Value} \\
\midrule
\midrule
\multicolumn{3}{c}{\textbf{\large Common to All Probes}} \\
\midrule
Hubble constant at present time & $H_0$ & $73.4$ \\
Present-day matter density parameter & $\Omega_m$ & $0.306$ \\
\midrule
\midrule
\multicolumn{3}{c}{\textbf{\large Additional for Supernovae (SN)}} \\
\midrule
Absolute magnitude of SNe Ia & $M_B$ & $-19.2435$ \\
Deviation from DDR (EM) - Standard/breaking & $\epsilon_0^{\mathrm{EM}}$ & $0/0.1$ \\
\midrule
\midrule
\multicolumn{3}{c}{\textbf{\large Additional for Gravitational Waves (GW)}} \\
\midrule
Deviation from DDR (GW) - Standard/breaking & $\epsilon_0^{\mathrm{GW}}$ & $0/0.1$ \\
\bottomrule
\end{tabularx}
\label{tab:fiducials}
\end{table*}

All the sampled parameters are assumed to have a uniform prior distribution informed by \cite{Planck18} results or \cite{Brout_2022}. We stress the fact that no Gaussian priors have been applied to $\Omega_bh^2$ for BAO data or to $M_B$ for SN data. This implies that when considering only BAO and SN data, without including GW, the sampling is performed on $r_d$ for BAO and by marginalizing over $M_B$ for SN. However, when GW are included, they provide independent information on $H_0$, allowing us to sample on $\Omega_bh^2$ for BAO while maintaining a flat prior, and to sample over both $M_B$ and $H_0$ for SN, always using uniform priors.

Concerning the DDR violation parameters, we work within two distinct assumptions:
\begin{itemize}
    \item Universal Violation Scenario (UV), where we assume that a single mechanism affects the two sectors in the same way, and therefore $\epsilon_{\rm EM}=\epsilon_{\rm GW}=\epsilon_0$;
    \item Independent Violation Scenario (IVS), where $\epsilon_{\rm EM}$ and $\epsilon_{\rm GW}$ are treated as independent.
\end{itemize}

\begin{table*}[h!]
\caption{\textbf{Top}: Prior distributions for the cosmological and models parameters described in \autoref{sec:models}. \textbf{Bottom}: List of free parameters for each observable combination under the Independent Violation Scenario (IVS) and Universal Violation Scenario (UVS) frameworks.}
  \centering
  \begin{tabularx}{\textwidth}{L C R}
    \hline\hline
    \multicolumn{2}{l}{\textbf{Parameters}}  & \textbf{Prior} \\
    \hline
    \multicolumn{3}{c}{\textbf{\quad\quad\quad\quad Cosmology}} \\
    \hline
    Present day matter content                      & $\Omega_m$     & $\mathcal U(0.01, 0.99)$   \\
    Present day baryonic matter content             & $\Omega_b h^2$ & $\mathcal U(0.015, 0.030)$ \\
    Present day Hubble constant                     & $H_0$          & $\mathcal U(20, 100)$      \\
    Absolute Magnitude                              & $M_B$          & $\mathcal U(-25, -15)$     \\
    Sound horizon scale                             & $r_d$          & $\mathcal U(100, 200)$     \\
    \hline\hline
    \multicolumn{3}{c}{\textbf{\quad\quad\quad\quad Independent Violation Scenario}} \\
    \hline
    Deviation from standard DDR (EM)                & $\epsilon_{\rm EM}$ & $\mathcal U(-0.5, 0.5)$ \\
    Deviation from standard DDR (GW)                & $\epsilon_{\rm GW}$ & $\mathcal U(-0.5, 0.5)$ \\
    \hline\hline
    \multicolumn{3}{c}{\textbf{\quad\quad\quad\quad Universal Violation Scenario}} \\
    \hline
    Deviation from standard DDR (both EM and GW)    & $\epsilon_0$        & $\mathcal U(-0.5, 0.5)$ \\
    \hline
  \end{tabularx}
  
  \centering
  \begin{tabularx}{\textwidth}{L C X}
    \hline\hline
    \textbf{Combination} 
      & \multicolumn{1}{c}{\textbf{Open parameters}} \\
    \cline{2-3}
    \hline
    \multirow{3}{*}{BAO + SN}
      & Cosmology & $\Omega_m,\,r_d$ \\
      & IVS       & $\epsilon_{\rm EM}$ \\
      & UVS       & $\epsilon_{0}$ \\
    \hline
    \multirow{3}{*}{BAO + GW}
      & Cosmology & $\Omega_m,\,\Omega_bh^2,\,H_0$ \\
      & IVS       & $\epsilon_{\rm GW}$ \\
      & UVS       & $\epsilon_{0}$ \\
    \hline
    \multirow{3}{*}{SN + GW}
      & Cosmology & $\Omega_m,\,M_B,\,H_0$ \\
      & IVS       & $\epsilon_{\rm EM},\,\epsilon_{\rm GW}$ \\
      & UVS       & $\epsilon_{0}$ \\
    \hline
    \multirow{3}{*}{BAO + SN + GW}
      & Cosmology & $\Omega_m,\,M_B,\,\Omega_bh^2,\,H_0$ \\
      & IVS       & $\epsilon_{\rm EM},\,\epsilon_{\rm GW}$ \\
      & UVS       & $\epsilon_{0}$ \\
    \hline\hline
  \end{tabularx}
  \label{tab:prior}
\end{table*}

\section{Results}\label{sec:results}

In this section, we present the main results of this work, focusing on the two different scenarios explored: the Universal Violation Scenario (UVS) and the Independent Violation Scenario (IVS).

\subsection{Universal Violation Scenario - UVS}

In the UVS, we assume that any potential deviations from the standard DDR, as described in Eq.~\eqref{eq:std_DDR}, are universal across both the electromagnetic and gravitational sectors. The sampled parameters and priors adopted for each combination of observables in this scenario are listed in \autoref{tab:prior} under the label UVS. This framework is explored under both the standard and the violating (broken) scenarios for the DDR.

\begin{figure*}[h!]
    \centering
    \begin{subfigure}[b]{0.48\textwidth}
        \centering
        \includegraphics[width=\textwidth]{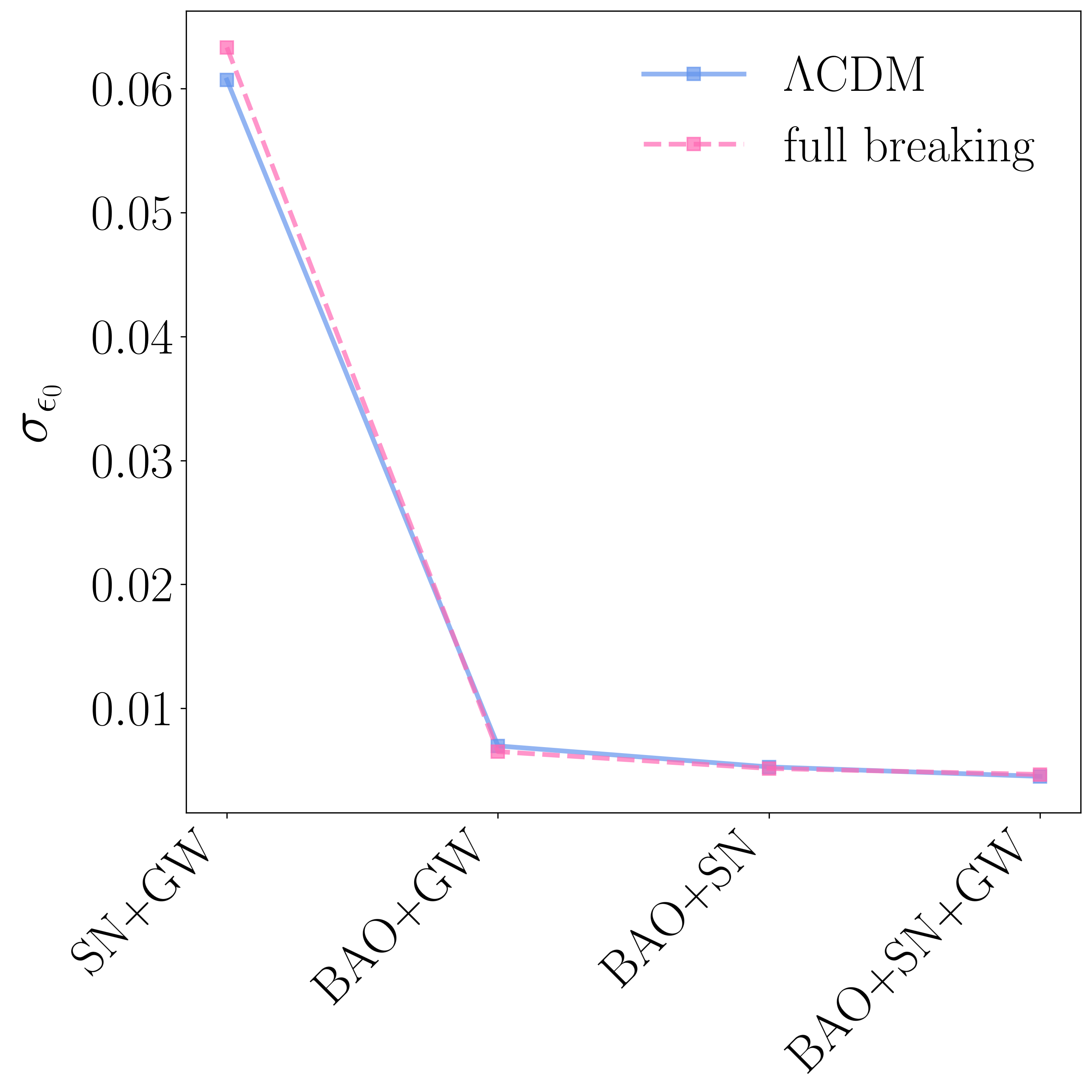}
        \caption{}
        \label{fig:epsilon0_sigma}
    \end{subfigure}
    \hfill
    \begin{subfigure}[b]{0.48\textwidth}
        \centering
        \includegraphics[width=\textwidth]{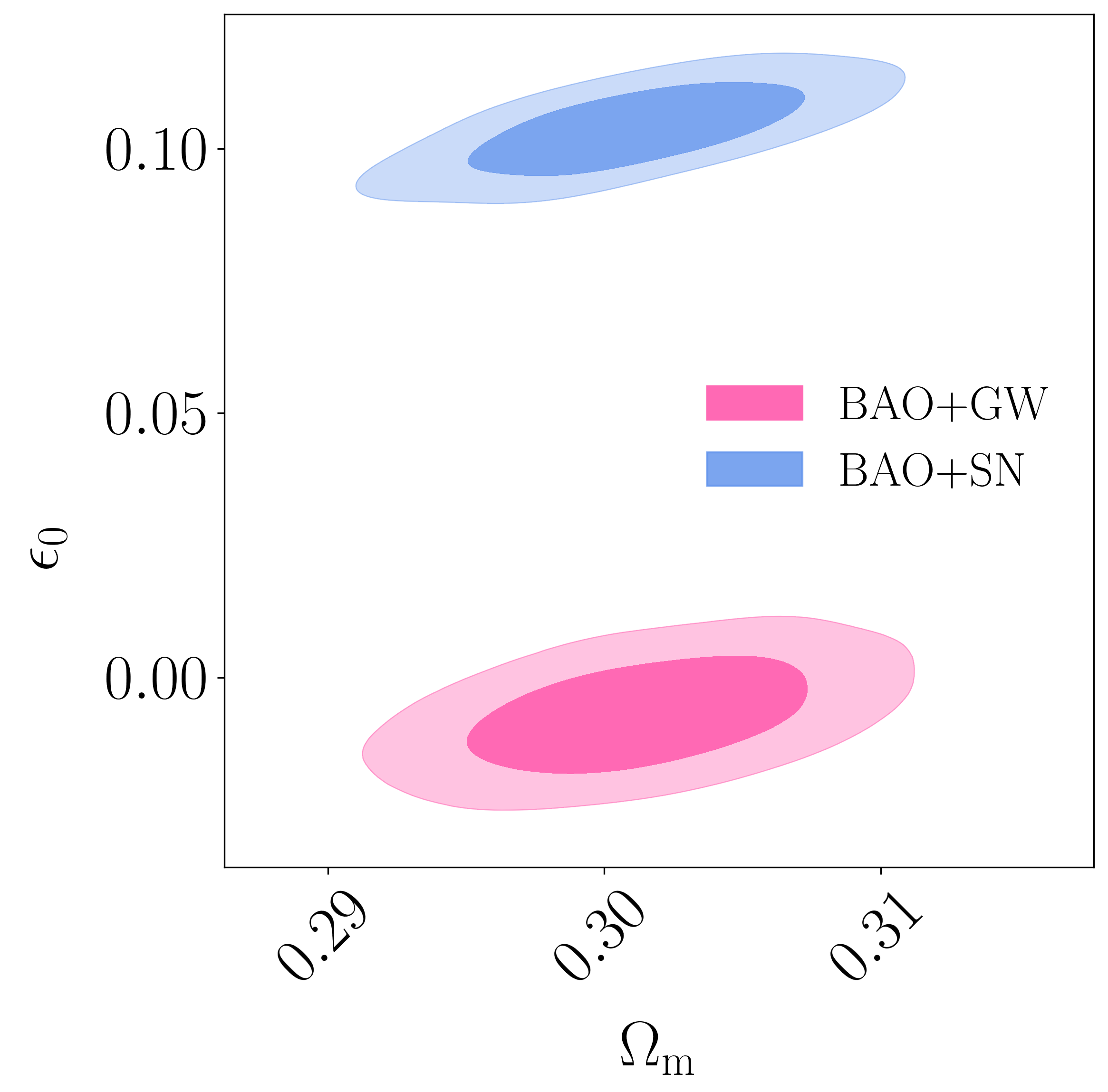}
        \caption{}
        \label{fig:epsilon0_omegam}
    \end{subfigure}
\caption{\textbf{(a) :} Error associated with the UVS parameter $\epsilon_0$ for different combinations of observables, shown for both the $\Lambda$CDM case (solid blue) and the breaking case (dashed pink).  \textbf{(b) : }Constraints in the $\epsilon_0$–$\Omega_m$ parameter space obtained assuming a universal violation scenario (UVS) affecting both the electromagnetic and gravitational sectors. If the true physical behavior instead involves a violation only in the electromagnetic sector, applying the UVS framework can artificially induce a tension in $\epsilon_0$ between the BAO+SN and BAO+GW datasets. }
    \label{fig:UVS_res}
\end{figure*}

\autoref{fig:epsilon0_sigma} shows the constraints on the UVS parameter $\epsilon_0$ for different combinations of observables. The role of BAO data is particularly significant: although BAO measurements are not sensitive to potential violations of the DDR, they constrain $\Omega_m$ effectively, thereby helping to break the degeneracy between $\Omega_m$ and $\epsilon_0$.
Furthermore, we observe that the uncertainty on $\epsilon_0$ remains relatively consistent between the standard and broken cases. In both scenarios, the combination of BAO+SN+GW yields the tightest constraint on $\epsilon_0$, though the inclusion of only SN or GW data individually does not significantly alter the associated uncertainty.

Another interesting result that highlights the impact of the inclusion of both EM and gravitational observables is shown in \autoref{fig:epsilon0_omegam}, where we present the constraints in the $(\epsilon_0, \Omega_m)$ parameter space obtained under the UVS assumption. 
We can notice that, if this is not the correct physical framework to describe the actual behaviour of the system (i.e., in the EM-breaking Universe), applying such an incorrect model can induce a tension in the UVS parameter $\epsilon_0$ between the BAO+SN and BAO+GW datasets. This outcome demonstrates how the use of an incorrect theoretical model can lead to misleading discrepancies between observational probes, emphasising the importance of employing a physically accurate framework when interpreting cosmological data. Furthermore, this highlights the complementarity of SN and GW, as the individual use of any of these two probes would not allow us to notice the inaccurate modelling of the physical effect.

The results concerning this case, both in the $\Lambda$CDM, EM-breaking and full-breaking Universes, are shown in \autoref{tab:UVS_results}.
\renewcommand{\arraystretch}{1.2} 
\begin{table*}[h!]
\centering
\caption{Constraints on $\Omega_m$ and $\epsilon_0$ for different dataset combinations, comparing the $\Lambda$CDM Universe and the full-breaking Universe within the UVS assumption. For the EM-breaking case, the joint analysis of SN+GW is not computed, since they are in tension.}
\label{tab:UVS_results}
\resizebox{\textwidth}{!}{
\begin{tabular}{|l|l|c|c|c|c|}
\hline
\textbf{Scenario} & \textbf{Parameter} & \textbf{SN + GW} & \textbf{BAO + GW} & \textbf{BAO + SN} & \textbf{BAO + SN + GW} \\
\hline
\multirow{2}{*}{UVS - $\Lambda$CDM} 
  & $\Omega_m$ & $0.267 \pm 0.065$ & $0.3012 \pm 0.0042$ & $0.3013 \pm 0.0041$ & $0.3012 \pm 0.0039$ \\
  & $\epsilon_0$ & $-0.037^{+0.069}_{-0.056}$ & $-0.0065 \pm 0.0070$ & $-0.0002 \pm 0.0053$ & $-0.0020 \pm 0.0045$ \\
\hline
\multirow{2}{*}{UVS - EM-breaking} 
  & $\Omega_m$ & -- & $0.3013 \pm 0.0040$ & $0.3011^{+0.0038}_{-0.0042}$ & -- \\
  & $\epsilon_0$ & -- & $-0.0070 \pm 0.0074$ & $0.1038 \pm 0.0059$ & -- \\
\hline
\multirow{2}{*}{UVS - Full-breaking} 
  & $\Omega_m$ & $0.329^{+0.084}_{-0.069}$ & $0.3014 \pm 0.0040$ & $0.3011 \pm 0.0040 $ & $0.3015 \pm 0.0040$ \\
  & $\epsilon_0$ & $0.113^{+0.077}_{-0.047}$ & $0.0946 \pm 0.0065$ & $0.0950 \pm 0.0051$ & $0.0950 \pm 0.0047$ \\
\hline
\end{tabular}}
\end{table*}
\renewcommand{\arraystretch}{1.0}

\subsection{Independent Violation Scenario}

In the IVS, we instead allow for deviations from the standard DDR to occur independently in the electromagnetic and gravitational sectors. Accordingly, the analysis introduces two separate parameters: $\epsilon_{\rm EM}$ and $\epsilon_{\rm GW}$, which capture possible violations in each sector. The sampled parameters and priors for this analysis are listed in \autoref{tab:prior}. We explore the three distinct cases for which we simulated the observations: the $\Lambda$CDM, full-breaking and EM-breaking Universes.

\begin{figure}[h!]
    \centering
    \includegraphics[width=0.9\linewidth]{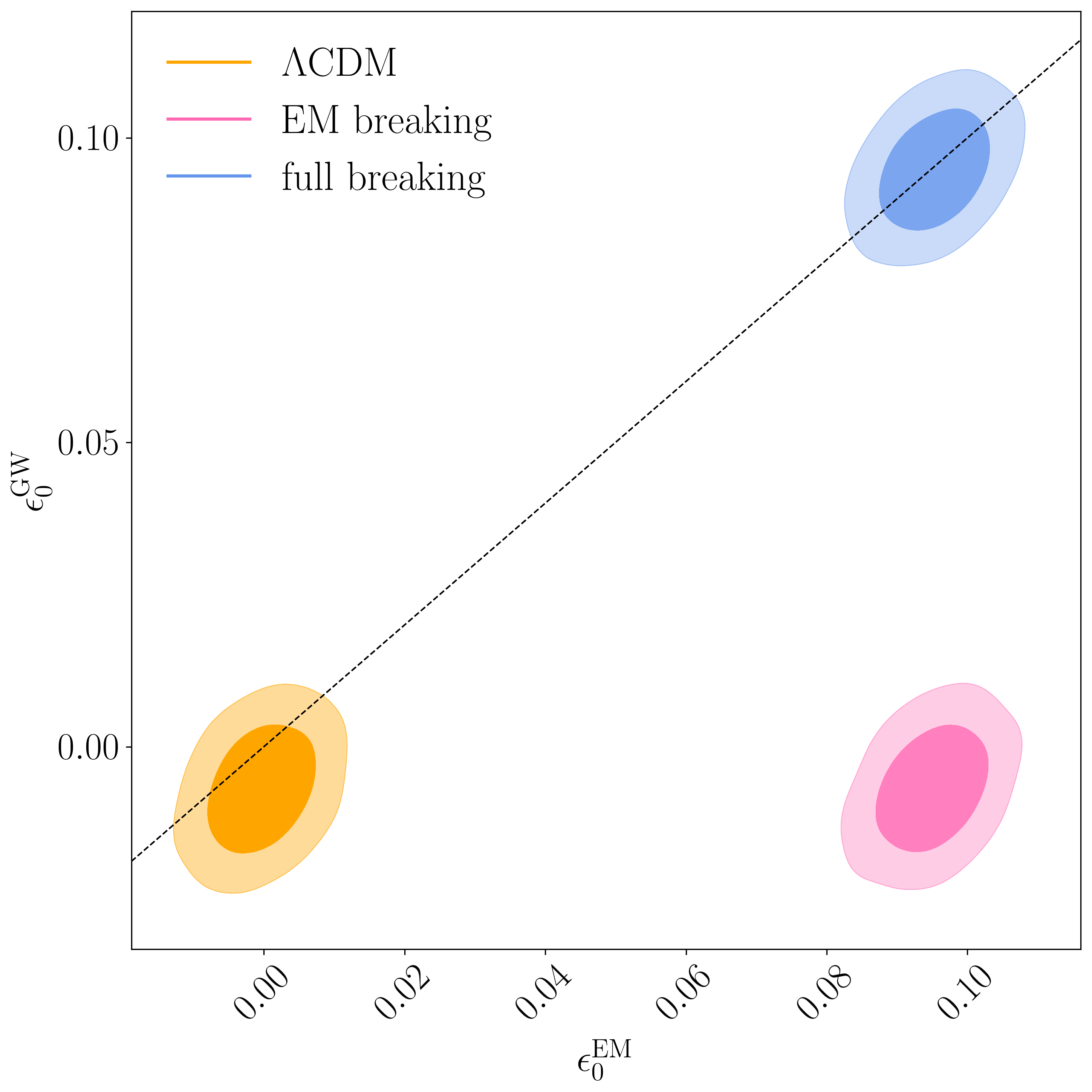}
    \caption{Constraints in the $\epsilon_0^{\rm EM}$–$\epsilon_0^{\rm GW}$ parameter space for three different physical scenarios: $\Lambda$CDM (orange), violation in the electromagnetic sector only (pink), and violation in both the electromagnetic and gravitational sectors (blue). The black dashed line represents the case $\epsilon_0^{\rm EM} = \epsilon_0^{\rm GW}$. }
    \label{fig:epsilonEMGW}
\end{figure}

These three cases are illustrated in \autoref{fig:epsilonEMGW}, and they yield clearly distinguishable patterns in the parameter space. This result highlights the power of combining SN and GW data with BAO measurements to effectively constrain deviations from the standard $\Lambda$CDM cosmology and the validity of the Etherington relation. Moreover, this approach allows us to distinguish the physical origin of such deviations. The addition of GW measurements is essential in this context. When the analysis is limited to the electromagnetic sector alone, the scenarios involving purely electromagnetic violations (pink) and combined electromagnetic and gravitational violations (blue) become indistinguishable, despite reflecting fundamentally different physics. Without access to the $\epsilon_0^{\rm GW}$ axis, the blue case would effectively collapse onto the pink one in the $\epsilon_{\rm EM}$–$\Omega_m$ parameter space, preventing us from distinguishing between the two scenarios. However, making this distinction is crucial, as they correspond to fundamentally different physical contexts.
In particular, the pink case, which involves a violation only in the electromagnetic sector, could point to mechanisms such as photon absorption by intergalactic dust or exotic particle physics processes. In contrast, the blue case suggests a violation that also affects the gravitational sector, potentially arising in modified gravity frameworks. A summary of the constraint obtained with this configuration in the three different physical scenarios can be seen in \autoref{tab:IVS_results}.

\renewcommand{\arraystretch}{1.2}
\begin{table*}[h!]
\centering
\caption{Constraints for the BAO+SN+GW dataset in the three different physical cases in the Independent Violation Scenario.}
\label{tab:independent_violation}
\begin{tabular}{|l|c|c|c|}
\hline
\multicolumn{4}{|c|}{BAO + SN + GW} \\
\hline
\textbf{IVS} & $\Omega_m$ & $\epsilon_0^\mathrm{EM}$ & $\epsilon_0^\mathrm{GW}$ \\
\hline
$\Lambda$CDM & $0.3012 \pm 0.0041$ & $-0.0004 \pm 0.0050$ & $-0.0068 \pm 0.0070$ \\
$\epsilon_{\mathrm{EM}}$ breaking & $0.3011 \pm 0.0040$ & $0.0950 \pm 0.0052$ & $-0.0068 \pm 0.0069$ \\
$\epsilon$ breaking & $0.3016 \pm 0.0040$ & $0.0954 \pm 0.0052$ & $0.0948 \pm 0.0066$ \\
\hline
\end{tabular}
\label{tab:IVS_results}
\end{table*}
\renewcommand{\arraystretch}{1.0}

\subsection{Padé parametrization}

The final results we present concern the Padé parametrization introduced in Eq.~\eqref{eq:eta_new_series}. This is a phenomenological approximation to hypothetical models in which the mechanism responsible for deviations from the standard DDR is naturally suppressed at high redshift. Such an approximation allows to avoid the modelling of early time physics in a context where DDR is violated, which could lead to complicated theoretical modelling. In this analysis, we explore the consequences of using such an approximation, when the DDR breaking mechanism is instead active at all redshifts.
\begin{figure}[h!]
    \centering
    \begin{subfigure}[b]{0.31\linewidth}
        \centering
        \includegraphics[width=\linewidth]{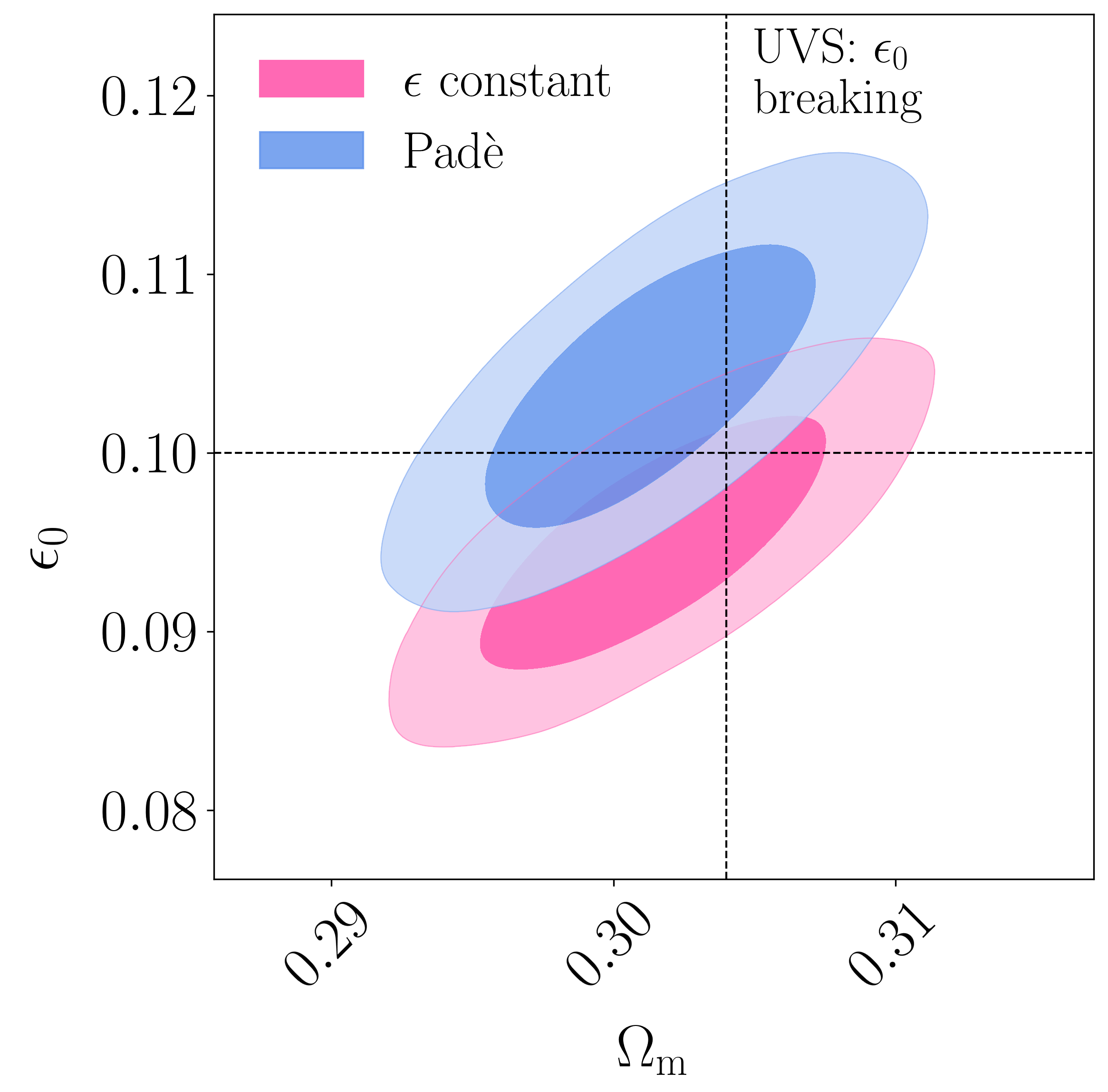}
        \caption{}
        \label{fig:epsilon0_epscost}
    \end{subfigure}
    \hfill
    \begin{subfigure}[b]{0.31\linewidth}
        \centering
        \includegraphics[width=\linewidth]{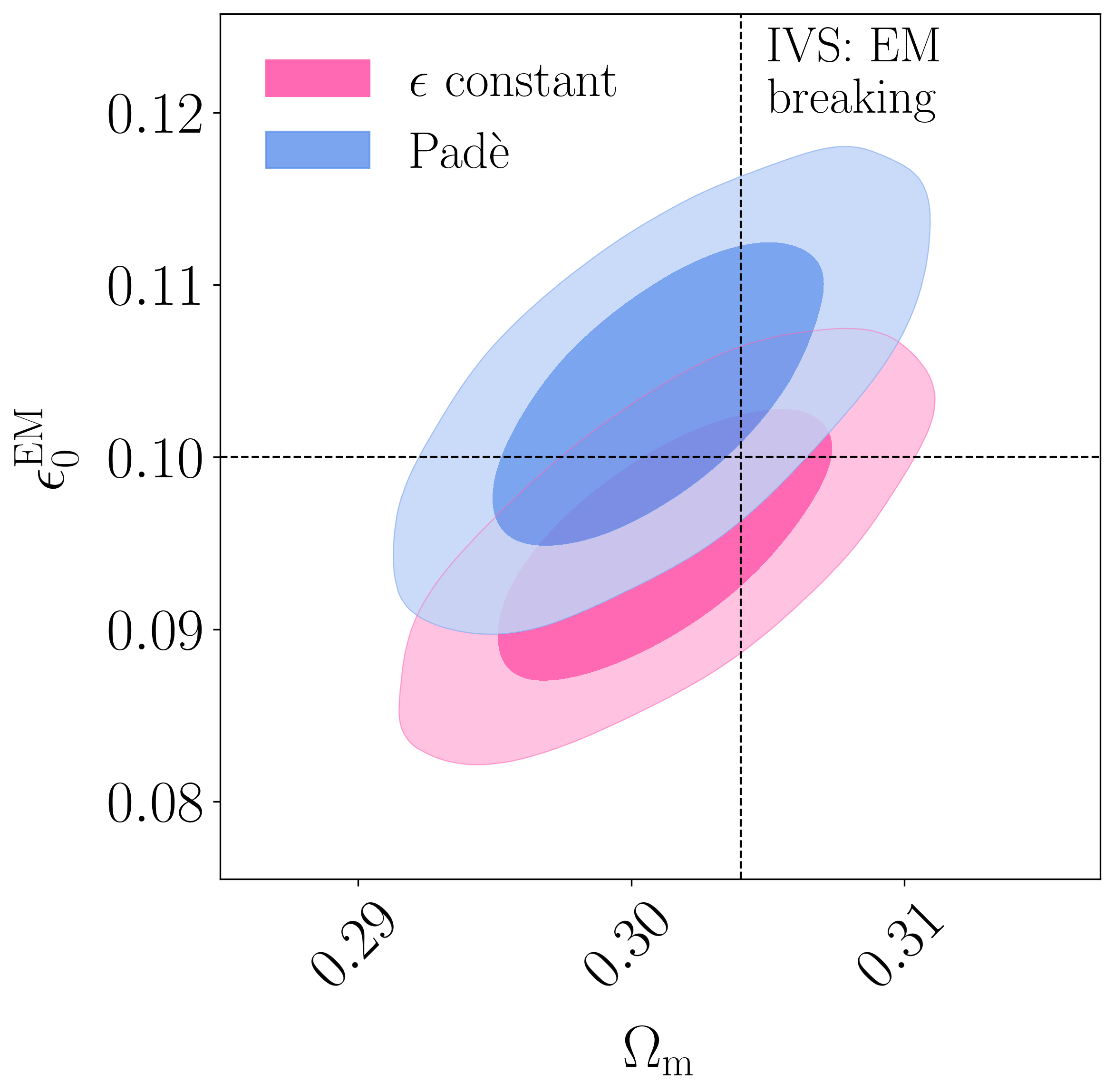}
        \caption{}
        \label{fig:epsilonEM_epscost}
    \end{subfigure}
    \hfill
    \begin{subfigure}[b]{0.31\linewidth}
        \centering
        \includegraphics[width=\linewidth]{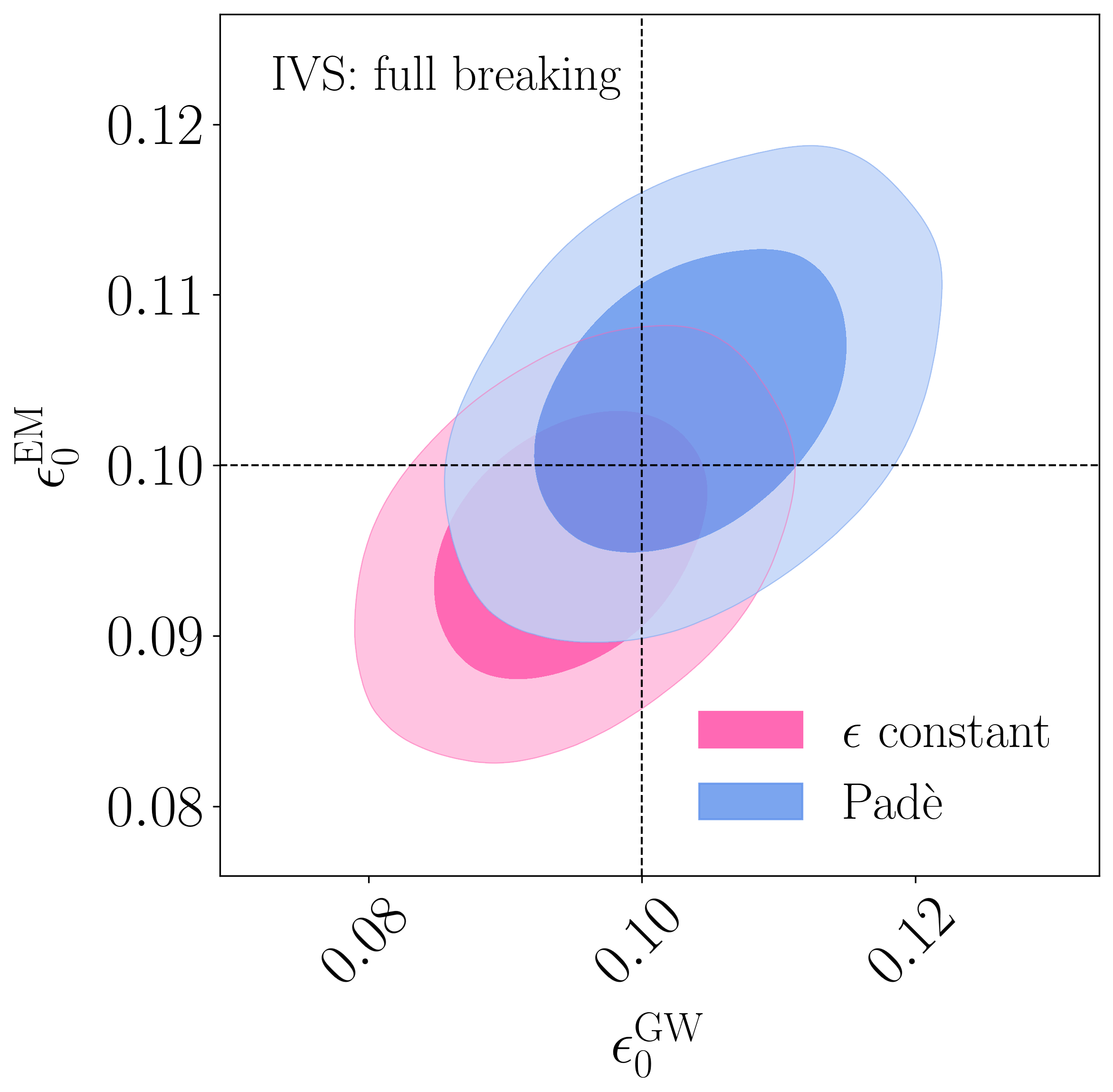}
        \caption{}
        \label{fig:epsilonEMGW_epscost}
    \end{subfigure}
    \caption{Two-dimensional posterior contours ($68\%$ and $95\%$ confidence levels) for the BAO+SN+GW joint analysis using mock data generated with the constant $\epsilon$ model (true model). Results are shown for two analysis assumptions: the correct constant $\epsilon$ model (pink) and the incorrect Padé parametrization (blue). The dashed lines indicate the fiducial values used in the data generation. The three breaking scenarios are shown: UVS (a), IVS EM-only (b), and IVS full (c).}
\label{fig:epscost_wrong}
\end{figure}

This comparison is performed within both the UVS and IVS frameworks by generating mock data using the constant $\epsilon$ parametrization defined in Eq.~\eqref{eq:eta0} and then analyzing this data assuming either the Padé or constant $\epsilon$ model. This is shown in \autoref{fig:epscost_wrong}. In each figure, we compare the same dataset analyzed under both assumptions for $\epsilon$, presenting only the two-dimensional contours from the joint BAO+SN+GW analysis.
Across all breaking scenarios, UVS, IVS-EM, and IVS-full, we observe a shift in the inferred DDR parameters when the wrong model is used. However, this shift is not statistically significant, as the results remain consistent within the 95\% confidence level. 
Consequently, we can conclude that the Pad\'e approximation can be safely used, at least for the redshift range of interest of the observables used in this work.

\renewcommand{\arraystretch}{1.8}

\begin{table*}[h!]
\centering
\resizebox{\textwidth}{!}{
\begin{tabular}{>{\centering\arraybackslash}m{4cm} |
                >{\centering\arraybackslash}m{3.5cm} |
                >{\centering\arraybackslash}m{3.5cm} |
                >{\centering\arraybackslash}m{3.5cm} |
                >{\centering\arraybackslash}m{3.5cm}}
    \toprule
    \textbf{Analysis Model} & $\Omega_m$ & $\epsilon_0$ & $\epsilon_0^{\rm EM}$ & $\epsilon_0^{\rm GW}$ \\
    \midrule
    UVS $\epsilon$ const & $0.3015\pm 0.0040$ & $0.0950\pm 0.0047$ & -- & -- \\
    UVS Padé             & $0.3011\pm 0.0041$ & $0.0892\pm 0.0048$ & -- & -- \\
    \midrule
    IVS-EM $\epsilon$ const & $0.3011\pm 0.0040$ & -- & $0.0950\pm 0.0052$ & $-0.0068\pm 0.0069$ \\
    IVS-EM Padé             & $0.3014\pm 0.0041$ & -- & $0.0927\pm 0.0052$ & $-0.0063\pm 0.0070$ \\
    \midrule
    IVS-EM+GW $\epsilon$ const & $0.3016\pm 0.0040$ & -- & $0.0954\pm 0.0052$ & $0.0948\pm 0.0066$ \\
    IVS-EM+GW Padé            & $0.3013\pm 0.0041$ & -- & $0.0924\pm 0.0052$ & $0.0789\pm 0.0078$ \\
    \bottomrule
\end{tabular}}
\caption{Constraints on DDR parameters from the BAO+SN+GW joint analysis. Results are categorized by physical scenario (UVS, IVS EM, IVS full) and the model used for $\epsilon(z)$ (constant or Padé).}
\label{tab:restructured_ddr_results}
\end{table*}

\section{Conclusions and outlook}\label{sec:conclusions}
In this work, we assess the sensitivity of future cosmological surveys to potential violations of the DDR, with a particular focus on the role of GWs as standard sirens. By combining luminosity distance measurements from SNe and GWs with angular diameter distance information from BAO, we explore the ability of future observations to distinguish between different physical mechanisms that could cause a breakdown of the DDR, either in the electromagnetic (EM) or gravitational sectors.

We consider two parametrizations for DDR breaking: a constant $\epsilon$ model and a Padé expansion, simulating mock datasets based on next-generation surveys (see \autoref{sec:data}). Our analysis covers two main theoretical frameworks: the UVS, where the EM and GW sectors are affected equally and described by a single parameter $\epsilon_0$, and the IVS, where violations are characterized by two distinct parameters, $\epsilon_0^{\rm EM}$ and $\epsilon_0^{\rm GW}$.

In the UVS case, we demonstrate that the inclusion of BAO significantly enhances constraints on the DDR parameter, reducing its uncertainty when combined with SN and GW data. We also show that, if the UVS assumption is incorrect, analyzing the EM and GW datasets separately leads to detectable tensions in the estimated $\epsilon_0$ values, offering a possible diagnostic tool to identify the class of physical models leading to the DDR violation.

Under the IVS framework, we analyze three scenarios: no violation, EM-only violation, and simultaneous EM and gravitational violations. We show that these cases lead to distinguishable signatures in parameter space. Crucially, we find that the inclusion of GW observations is essential to discriminate between violations originating in the EM sector, such as photon number non-conservation from intergalactic dust or exotic particles, and those extending to the gravitational sector, potentially linked to modified gravity theories.

Finally, we investigate the implications of commonly used phenomenological approximations
by comparing results obtained when analysing datasets with the incorrect DDR-breaking parametrization. We compare scenarios where the true underlying physics is well described by the constant $\epsilon$ parametrization and find that, in all cases, using the wrong model introduces a shift in the estimated DDR parameters. However, this shift remains within the $2\sigma$ level, indicating the need for a more extended redshift range for the observations, if one wants to distinguish between these cases.

A promising direction for future work is to explore specific modified gravity models or physical mechanisms that naturally lead to a Padé-like redshift dependence in DDR violations. Incorporating such models into the analysis would enable a more meaningful interpretation of the observational constraints and provide a clearer assessment of how gravitational wave data enhance our ability to differentiate between alternative physical explanations for DDR violations.

\section*{CRedIT statement} 

\textbf{Chiara De Leo}: Conceptualization; Methodology; Software; Visualization; Writing – original draft; Writing – review \& editing;
\textbf{Matteo Martinelli}: Conceptualization; Methodology; Software; Writing – original draft; Writing – review \& editing;
\textbf{Rocco D'Agostino}: Conceptualization; Methodology; Writing – original draft; Writing – review \& editing;
\textbf{Giulia Gianfagna}: Methodology; Software; Writing – original draft; 
\textbf{Carlos J. A. P. Martins}: Conceptualization; Methodology; Writing – review \& editing.

\acknowledgments

CDL thanks iniziativa specifica INFN for financial support. This work
was partially supported by the research grant number 2022E2J4RK "PANTHEON:
Perspectives in Astroparticle and Neutrino THEory with Old and New messengers"
under the program PRIN 2022 funded by the Italian Ministero dell’Universit\`a
e della Ricerca (MUR). MM acknowledges funding by the Agenzia Spaziale Italiana (\textsc{asi}) under agreement n. 2024-10-HH.0 and support from INFN/Euclid Sezione di Roma. 
RD acknowledges financial support from INFN -- Sezione di Roma 1, \textit{esperimento} Euclid,  and work from COST Action CA21136 -- Addressing observational tensions in cosmology with systematics and fundamental physics (CosmoVerse) supported by European Cooperation in Science and Technology.
CDL and MM acknowledge financial support from Sapienza Università di Roma, provided through Progetti Medi 2021 (Grant No. RM12117A51D5269B). This work made use of Melodie, a computing infrastructure funded by the same project, and PLEIADI, a computing infrastructure installed and managed by INAF.
\newline
This work was financed by Portuguese funds through FCT (Funda\c c\~ao para a Ci\^encia e a Tecnologia) in the framework of the project 2022.04048.PTDC (Phi in the Sky, DOI 10.54499/2022.04048.PTDC). CJM also acknowledges FCT and POCH/FSE (EC) support through Investigador FCT Contract 2021.01214.CEECIND/CP1658/CT0001 \newline
(DOI 10.54499/2021.01214.CEECIND/CP1658/CT0001).

\bibliographystyle{JHEP}
\bibliography{biblio.bib}

\end{document}